\documentclass[12pt]{spieman}  
\usepackage{amsmath,amsfonts,amssymb}
\usepackage{graphicx}
\usepackage{setspace}
\usepackage{tocloft}
\usepackage{lineno}
\usepackage[final]{changes}

\title{The Engineering Development Array 2: design, performance and lessons from an SKA-Low prototype station}

\author[a]{Randall~Wayth}
\author[a]{Marcin~Sokolowski}
\author[a]{Jess~Broderick}
\author[a]{Steven~J.~Tingay}
\author[a]{Raunaq Bhushan}
\author[a]{Tom~Booler}
\author[b]{Riccardo Chiello}
\author[a]{David~B.~Davidson}
\author[a]{David Emrich}
\author[a]{Budi Juswardy}
\author[a]{David Kenney}
\author[c]{Giulia Macario}
\author[d]{Alessio Magro}
\author[e]{Andrea Mattana}
\author[a]{David Minchin}
\author[e]{Jader Monari}
\author[a]{Andrew McPhail}
\author[e]{Federico Perini}
\author[e]{Giuseppe Pupillo}
\author[e]{Marco Schiaffino}
\author[f]{Ravi Subrahmanyan}
\author[g]{Andre van Es}
\author[a]{Mia Walker}
\author[g]{Mark Waterson}

\affil[a]{ICRAR/Curtin University, Curtin Institute of Radio Astronomy, Kent St, Bentley, Australia, 6102}
\affil[b]{University of Oxford, Department of Physics, Denys Wilkinson Building, Oxford OX1 3RH, UK}
\affil[c]{Istituto Nazionale di Astrofisica (INAF), Osservatorio Astrofisico di Arcetri, Largo Enrico Fermi 5,  Firenze, 50125, Italy}
\affil[d]{Institute of Space Sciences and Astronomy, University of Malta, Msida, MSD2080, Malta.}
\affil[e]{Istituto di Radioastronomia, Istituto Nazionale di Astrofisica  - Via Gobetti 101, 40129 Bologna ITALY}
\affil[f]{Space \& Astronomy, Commonwealth Scientific and Industrial Research Organisation, 26 Dick Perry Avenue, Kensington WA 6151, Australia}
\affil[g]{SKA Observatory, Jodrell Bank, Lower Withington, Macclesfield, SK11 9FT, UK}

\cftpagenumbersoff{figure}
\cftpagenumbersoff{table} 
\begin{document} 
\maketitle

\begin{abstract}
We present the Engineering Development Array 2, which is one of two instruments built as a second generation prototype station for the future Square Kilometre Array Low Frequency Array. The array is comprised of 256 dual-polarization dipole antennas that can work as a phased array or as a standalone interferometer. We describe the design of the array and the details of design changes from previous generation instruments, as well as the motivation for the changes.
Using the array as an imaging interferometer, we measure the sensitivity of the array at five frequencies ranging from 70 to 320\,MHz.
\end{abstract}


{\noindent \footnotesize\textbf{*}R. Wayth,  \linkable{r.wayth@curtin.edu.au} }

\begin{spacing}{1.25}

\section{Introduction}
\label{sect:intro} 
The next generation of radio telescopes, designed to explore the history and evolution of the Universe \cite{Braun:2015B3}, will be large scale facilities in order to achieve high sensitivity and high angular resolution, in some cases over large fractional bandwidths.  The key science missions for these facilities also demand extraordinary measurement accuracy and precision, requiring stringent control over the calibration of these measurements \cite{2019cosm.book.....M}.
This will be achieved by establishing a comprehensive understanding of the telescope system and reflecting that understanding in the design and manufacture of the physical signal chain and the calibration and data processing software.

In the case of the low frequency component of the Square Kilometre Array \cite{SKA1L1req}, SKA-Low, aperture array technologies are used to achieve high sensitivity through the use of very large numbers of conceptually simple individual antennas.  \replaced{The current design of phase 1 SKA-Low (SKA1-Low) has approximately}{The current design of SKA-Low has} 130,000 individual antennas arranged as 512 `stations', each composed of 256 dual-polarised antennas \added{working as one or more phased arrays}\cite{SKABaselinev2}. SKA-Low continues a development path established by LOFAR \cite{2013A&A...556A...2V},
the Murchison Widefield Array (MWA) \cite{2009IEEEP..97.1497L,2013PASA...30....7T,2018PASA...35...33W},
and PAPER \cite{2010AJ....139.1468P} as SKA-Low precursors and pathfinders.
When fully deployed at the Murchison Radio-astronomy Observatory (MRO) later this decade, SKA1-Low will be the most sensitive telescope ever built in its frequency range (50-350\,MHz).

The design and requirements specification process for SKA-Low has been a long one, moving through many stages of antenna selection, subsystem design, unit development and prototyping, and small-scale deployments to the MRO, culminating in the deployment of a full-scale prototype station called Aperture Array Verification System 1 (AAVS1) in 2016 and 2017 \cite{2016icea.confE...1H,2017gass.confE...1B,BenthemAAVS1}.
In the time leading up to AAVS1, the baseline design for SKA1-Low used the SKALA2\cite{7297231} antenna. In addition, as part of an SKA cost control process in 2015-2016, another prototype station using MWA-style dipoles and a hybrid beamforming architecture, called the Engineering Development Array\cite{2017PASA...34...34W} (EDA) was also deployed.


The deployment and commissioning process for AAVS1 generated a long list of lessons-learned\cite{BenthemAAVS1}.
Arguably the most important outcome from AAVS1 was a full appreciation of the importance and effects of mutual electromagnetic coupling within the antennas in the station.
Post AAVS1, the issue of intra-station mutual coupling, and the effects on the station calibratability and beam stability, was seen as a significant risk such that two successor stations to AAVS1 were built: AAVS2\cite{2020SPIE11445E..89V} and EDA2.
Both of these next-generation prototypes incorporated many improvements (described in this paper); in addition AAVS2 uses an updated SKALA antenna: the SKALA4.1AL\cite{9107113}. 
After AAVS1, a substantial amount of effort has gone into modelling and understanding the computational, calibration and scientific effects of mutual coupling, especially in the SKALA family of antennas,\cite{8739902,8879294,2020Steiner,9232307,2021MNRAS.tmp.1387J,Bolli2021} and verifying their performance with drone\cite{9232190} and astronomical\cite{9410962,Macario2021} measurements.

Given the myriad challenges noted above, a critical component of this prototyping program is the development of a well-understood and trusted comparator system, the EDA2.
The requirements on this comparator system were that it have an identical overall architecture to the station composed of SKALAs, but utilising a well-understood individual antenna element.  The individual antenna should be inexpensive, have a sensitivity that meets the SKA-Low requirements specification in the frequency range that supports the most critical SKA-Low science (100 - 200 MHz), and be electromagnetically simpler than the SKALA antenna (in order to avoid the mutual coupling effects).  The intent of the comparator system is to decouple the general considerations of the overall system development from the additional complexities inherent in the SKALA implementation.  The comparator system also allows a more straightforward comparison between measurement and simulation, providing confidence in both the simulation results and the measurement system.

The antenna chosen that meets the requirements of the comparator system is the so-called bowtie dual-polaristion dipole antenna utilised for the MWA, which has a common heritage with the LOFAR high-band antenna.  This technology has been in operation as part of LOFAR since 2012 and as part of the MWA since 2013.  The antenna has been well tested and characterized in the field at a high level of detail \cite{2015RaSc...50...52S,2015RaSc...50..614N,2017PASA...34...62S,2018PASA...35...45L,9040892,2021MNRAS.502.1990C}.

The resulting comparator system is called the Engineering Development Array \#2 (EDA2).  As for the SKALA prototype stations, two iterations of the EDA\cite{2017PASA...34...34W} have been developed.
In addition to the verification and system commissioning work described in this paper, EDA2 has also been used for targetted science programs\cite{2020MNRAS.499...52M,2021PASA...38...23S}.

We describe below the second generation engineering development array, EDA2.
In section \ref{sec:design_const} we describe the design and construction and illustrate new components and architectural changes used in EDA2.
In section \ref{sec:datacal} we describe the observation, calibration and data analysis techniques used.
In section \ref{sec:perf} we present measurements of the system sensitivity, compare to SKA specifications, and discuss limitations of the current system.
We conclude and discuss the future work in section \ref{sec:conclusion}.

\section{Design and Construction}
\label{sec:design_const}
\begin{figure}
\includegraphics[width=\linewidth]{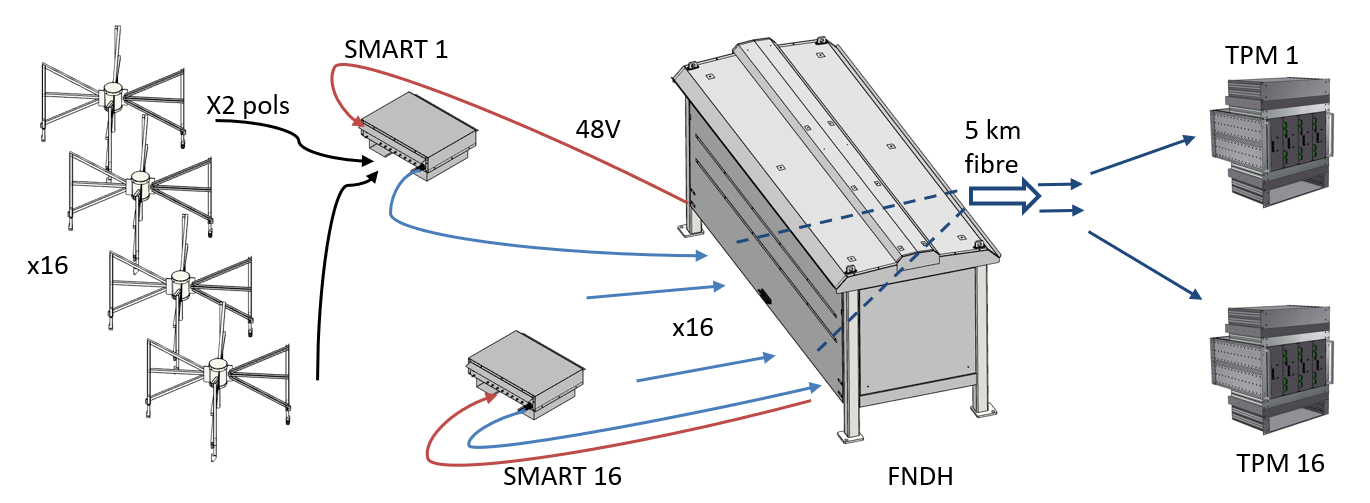}
\caption{EDA2 block diagram. Groups of 16 antennas are connected to a SMART box. Each SMART box is connected to the nearby FNDH, which patches the optical signals to long distance fibre connection to the MRO control room. There, the fibres are connected in groups of 16 to TPMs.}
\label{fig:eda2_diagram}
\end{figure}

The EDA2 comprises 256 dual-polarization dipoles pseudo-randomly distributed over a diameter of 35\,m. A block diagram of the EDA2 layout and signal path is shown in Fig \ref{fig:eda2_diagram}.
The layout of EDA2 is intentionally identical to that of the EDA1\cite{2017PASA...34...34W} and AAVS1\cite{BenthemAAVS1} (with the exception of the central antenna in AAVS1) so that direct comparisons of the systems can be made. \added{For the same reason, AAVS2 uses the same layout, but is scaled up by approximately 10\%.\cite{Macario2021}}

The dipoles used in EDA2 are identical to EDA1: they are MWA-style dual-polarization dipoles whose low-noise amplifiers (LNAs) are modified to respond down to 50\,MHz. Dipoles are attached to a wire ground mesh with square size 50\,mm, filled out to a minimum diameter of 40\,m.
The mesh is aligned with local cardinal directions; hence the dipole arms are aligned with North-South and East-West.
The dipoles are connected to ``SMART'' boxes, which are described in detail below, via 7\,m of KSR100 coaxial cable. Each SMART box aggregates 16 dipoles, converts the signals from electrical to optical, and sends signals downstream as RF-over-Fibre (RFoF) on 24-core loose-tube optical fibre cables.

Signals from each SMART box in EDA2 are aggregated in a nearby Field Node Distribution Hub (FNDH), where optical fibres are connected to long distance (approximately 5\,km) optical fibre connections to the Observatory control room. The FNDH also provides and controls power to SMART boxes.

In the control room, optical signals are patched through to 16 Tile Processing Modules\cite{2017JAI.....641014N} (TPMs). The EDA2 uses identical TPMs to AAVS1 and AAVS2 and shares a common suite of software for monitor \& control, data capture and data processing.

\subsection{SMART boxes}
A significant change to the signal path from AAVS1 was the introduction of ``Small Modular Aggregation \& RFoF Trunk'' (SMART) boxes. As their name suggests, each SMART box provides an aggregation point for the signals from 16 antennas, and converts the signals to optical RFoF for transmission to the Control Building. The SMART box also provides a DC bias to power the dipoles via their coaxial cables.

Motivation for creation of the SMART boxes came from the experiences of AAVS1, in particular:
\begin{itemize}
\item The single aggregation point in the center of AAVS1 created an unwieldy tangle of cables.
\item The bespoke hybrid cables that carried fibre and power to the apex of each antenna in AAVS1 were costly and impractical.
\item The ``trumpet'' component of AAVS1 at the apex of each antenna combined an LNA and RFoF module into a single small container and made connection and disconnection of the hybrid cable impractical.
\item The combination of issues listed above made maintenance work or debugging of the antennas and station very time and labour intensive.
\end{itemize}

The primary function of each SMART box is to connect antennas to the optical Front End Modules (FEMs), which convert the electrical RF signals to optical RFoF. The FEMs are described in detail in a companion paper\cite{Perini2021}.
The SMART box comprises a shielded power supply, which generates 5\,V from the 48\,V supplied from the FNDH.
Each FEM has two RF inputs (one for each polarization of a dipole) and provides 5\,V DC bias to power the dipoles via the coaxial cable. Internally, the FEM provides some amplification then converts the signal from electrical to optical via intensity modulation of two lasers working at different wavelengths, one for each polarization. Thus the FEM uses wavelength division multiplexing to combine the optical signals from both polarizations as different optical wavelengths onto a single output fibre.
The 16 output fibres are connected to the 24 core cable, which runs to the FNDH, with some spare fibres in case of breakage etc.

\begin{figure}
\includegraphics[width=\linewidth]{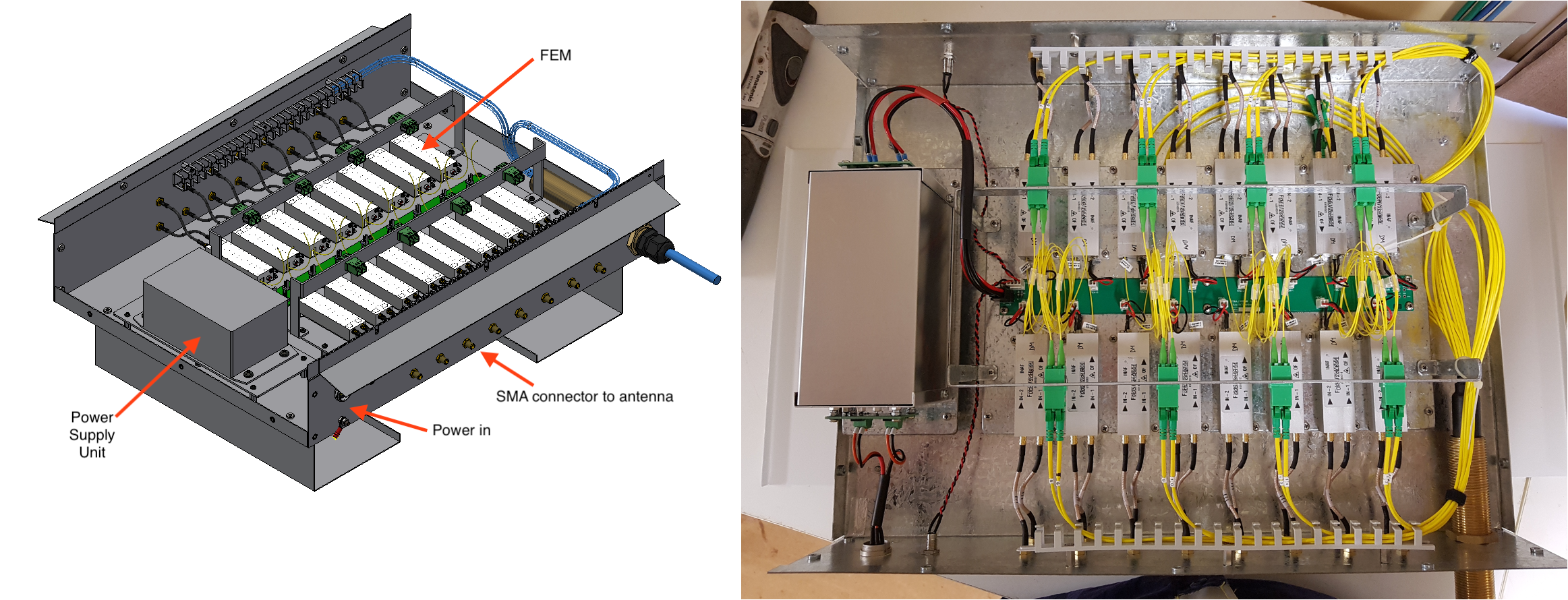}
\caption{Left: Annotated CAD drawing of a SMART box. Right: inside an assembled SMART box.}
\label{fig:SMART_inside}
\end{figure}

The same SMART box design is used for AAVS2\cite{2020SPIE11445E..89V}. Early in the deployment phase of EDA2 and AAVS2, an improvement was made to the FEMs to eliminate low-level low-frequency spurious emission from the optical modules\cite{9232422,2020JLwT...38.5393N} by introducing a very low frequency ($\sim 10$\,kHz) dithering signal to the laser bias. This modification was successful in eliminating the spurious emission and has no impact on the performance of the system.

\subsection{Layout}

\begin{figure}
\includegraphics[width=\linewidth]{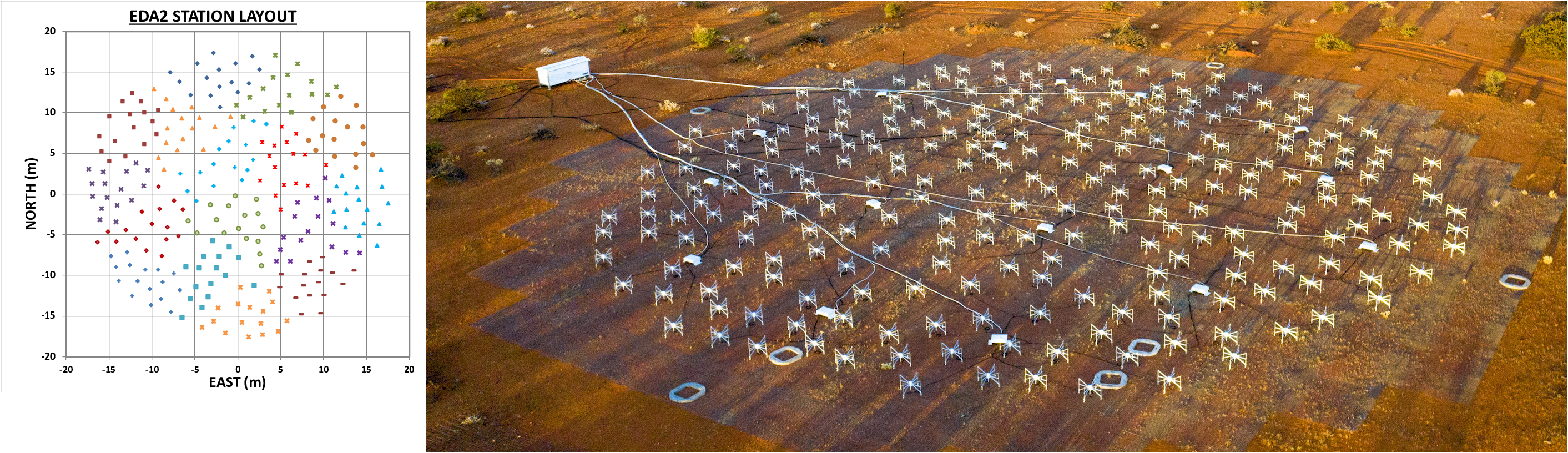}
\caption{Left: The antenna layout of the EDA2, with symbols according to SMART box connection. Right: Aerial view of the EDA2, looking northeast. The small white boxes in the array are SMART boxes. The large white box outside the array at the top left of the image is the FNDH.}
\label{fig:eda2_layout}
\end{figure}

The layout of EDA2 is shown in Fig. \ref{fig:eda2_layout} with symbols showing the mapping between dipoles and SMART boxes, as well as an aerial view of the completed array.

EDA2 was deployed in phases beginning with the ground mesh in Nov 2018.
Dipoles were assembled and installed onto the ground mesh without LNAs, after surveying for accurate locations, in Feb 2019. The SMART boxes were assembled by a joint Curtin/INAF team in April 2019 to be deployed for both EDA2 and AAVS2.
In April the FNDH enclosure was installed and temporary power and signal distribution arrangements were established, for both AAVS2 and EDA2, that enabled 48 antennas on each array (3 working SMART boxes per array), for initial testing and verification.
The remaining dipole LNAs and SMART boxes were installed in May 2019. Finally in Jul/Aug 2019, the FNDH power supply unit was installed and final fibre arrangement implemented, completing the infrastructure required to operate the full array.

\subsection{Signal levels in the EDA2 analogue chain}
\label{sec:siglevels}
The MWA dipoles used for EDA2 are sensitive to frequencies from 50 to 350\,MHz, however a significant difference between the MWA dipole and the SKALA antennas is the gain in the antennas themselves. For the SKALAs, the gain is approximately 40\,dB, whereas the MWA dipoles have approximately 20\,dB. This has some downstream consequences as discussed below, and we note here that the decision was made early in the planning for EDA2 to use the MWA dipoles as-is, rather than incur the risk and delay associated with redesigning the electronics inside the MWA dipole.

The FEM and downstream signal chain was designed for the higher gain SKALA antennas. The ``preADU'' modules in the TPMs provide filtering and amplification and have an adjustable gain range of 31\,dB\cite{BenthemAAVS1}, which is sufficiently large to be able to equalize the signal power from all antennas given the natural variation in optical gain of the FEMs.
The overall signal budget in the system was thus designed so that the typical gain adjustment in the TPMs is in the middle of its range, around 15\,dB when connected to the higher gain SKALA LNAs.

Based on the lower gain of the MWA dipoles LNAs, the preADUs used for the EDA2 were set at their highest gain, although in some cases this is not enough to equalize the digitised power for all signals. This will cause a small increase quantization noise for some dipoles, however we expect this to be negligible.

The second consequence of using the lower gain dipole is on the system noise temperature. The FEMs have a substantial noise figure (approximately 10\,dB), which must be hidden by high gain in the electronics upstream of it. The lower gain of the MWA dipoles cannot hide all of this noise, which artificially raises the system temperature of the EDA2 compared to the actual noise temperature of the MWA dipoles, which ranges from 40-100\,K over most frequencies\cite{9040892}.
We discuss this in some detail in Section \ref{sec:perf}.

\section{Data processing and Calibration}
\label{sec:datacal}
The core data capture and processing tasks used for EDA2 are very similar to those already developed for AAVS1\cite{BenthemAAVS1}, which have subsequently been used for an analysis of AAVS2\cite{9410962,Macario2021} sensitivity and for EDA2 transient science\cite{2021PASA...38...23S}.
As such, we give only a brief overview here and refer the reader to prior work for details.

Signals from each antenna are digitised within the TPMs at 800\,Msamp/sec then channelized into 512 coarse channels of width 0.925926\, MHz using an oversampled polyphase filterbank with oversampling ratio 32/27.
Coarse channels are separated by 0.781250\,MHz.
Static delays (due to cable length differences) are corrected in the TPMs before channelization, such that all signals are time-aligned to within a sample. The remaining residual delay (and any desired beam steering) is corrected via a phase applied to each coarse channel.
\begin{figure}
\includegraphics[width=\linewidth]{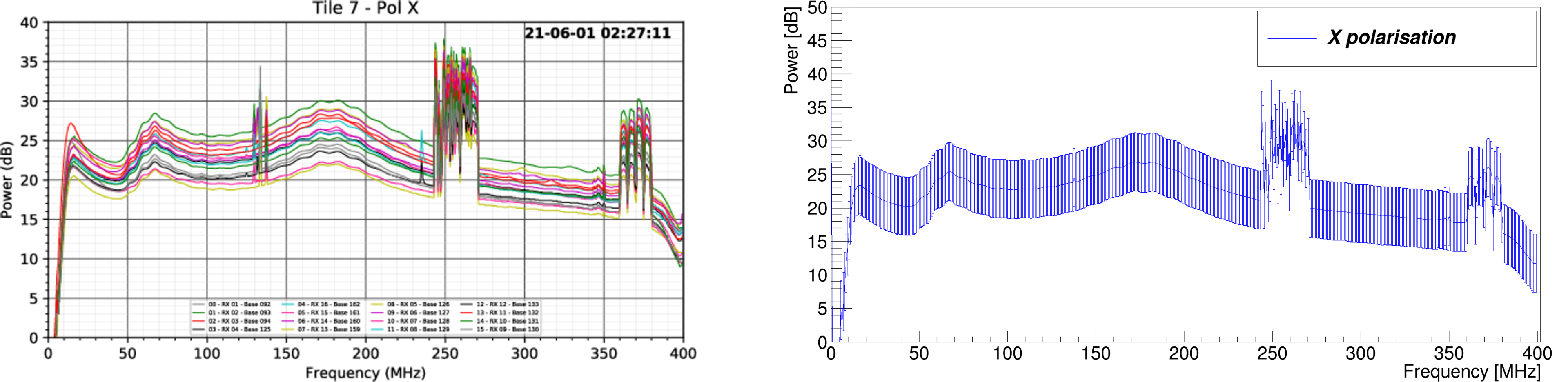}
\caption{Example spectra from a TPM/SMART box in EDA2. Left: total power spectra for each X polarization antenna. The image includes intermittent RFI around 130-140\,MHz due to aircraft communications and satellite downlinks. Persistent RFI is seen between 245-270 and 360-380\,MHz. Right: the X-polarization average spectrum over all dipoles with the $2\sigma$ range of power levels. The Y polarizations look very similar.}
\label{fig:example_spectra}
\end{figure}

\subsection{Data products}
The TPMs continuously produce total-power spectra for each antenna, which can be used for diagnostic and system monitoring purposes. An example of `X' polarization (east-west oriented dipoles) spectra for one TPM is shown in Fig. \ref{fig:example_spectra}. As noted above, the lower gain of the MWA LNAs cannot be fully compensated for in the TPMs, so there is a spread of signal powers, however the overall shape is the same for each antenna. Also shown in Fig. \ref{fig:example_spectra} is the average spectrum across all 256 antennas with the $2\sigma$ spread of power around the mean. The images also contain examples of intermittent \added{(aircraft and satellite around 140\,MHz)} and persistent \added{(satellite 245-270\,MHz and 360-380\,MHz)} radio-frequency interference (RFI), which we avoid for commissioning and system performance measurement.

The TPMs can produce three main data outputs: 1) the station voltage beam formed by adding phased voltages from all antennas; 2)  continuously streaming visibilities for a single coarse channel formed by cross-correlating all antennas within the station, with approximately 2\,s time resolution; 3) short (approximately 0.28\,s) dumps of voltages from all antennas for a coarse channel, which can be done once every few minutes.

In this paper, we use datasets taken in voltage-dump mode in April 2020 at five ``spot frequencies'', which are specified by the SKA for verification of the prototype arrays. Data were taken to cover at least the following frequencies: 70.3, 110.2, 159.4, 229.7 and 320.3\,MHz. Voltage-dump mode allows us to offline correlate the data and form two snapshots of the sky separated by 0.14\,s, as described in Sect. \ref{sec:perf}. Data used to present the calibration phase solutions was taken on 2020-04-10 in streaming correlator mode at 159\,MHz.

\subsection{Calibration}
\label{sec:pha_cal}

EDA2 was calibrated with the same procedure used for AAVS1\cite{BenthemAAVS1}. The sun is used over most frequencies as a strong compact calibration source with known flux density\cite{2009LanB...4B..103B} when in the quiet state. For the duration of the EDA2 commissioning and verification observations, the sun was indeed very quiet. Baselines shorter than $3-5\lambda$ (depending on frequency) were not used in calibration, however the data were deliberately taken in April when the Galactic plane is below the horizon around midday, so that the data will be least affected by any large-scale emission.

\begin{figure}
\includegraphics[width=\linewidth]{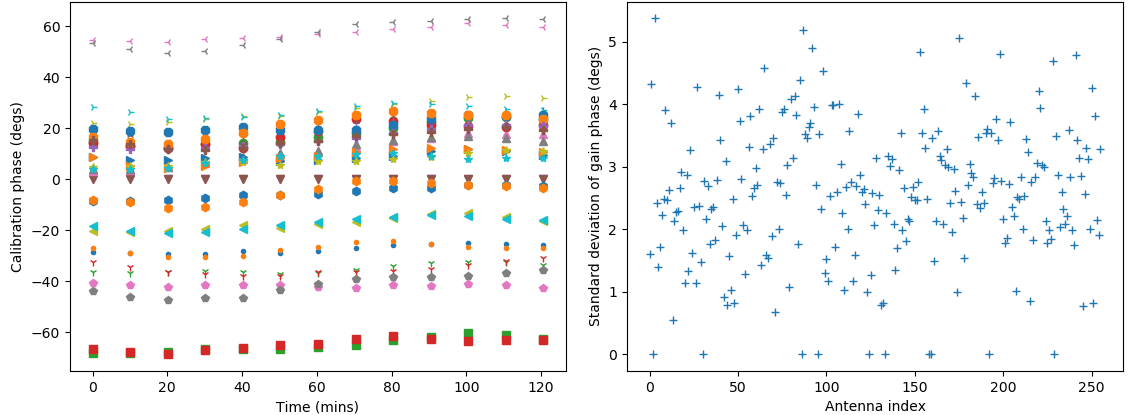}
\caption{Example calibration phase stability for sun-based calibration solutions at 159\,MHz, computed every 10 minutes over a 2 hour window centered on solar transit. Left: examples of calibration phase for the first 16 antennas (X and Y polarizations are plotted and generally lie on top of each other). Right: the standard deviation of the calibration phase for all antennas. Antennas with zero standard deviation were flagged.}
\label{fig:cal_pha_example}
\end{figure}

The calibration phase solutions for antennas were found to be very stable over hour and day-long timescales, similar to previous results. Example calibration phases (which are relative to an arbitrary phase reference antenna) are shown in Fig. \ref{fig:cal_pha_example} for 2 hours of data taken at 159\,MHz, centered around the solar transit.
Also shown in Fig. \ref{fig:cal_pha_example} is the standard deviation of the calibration phase calculated over all antennas. This figure clearly shows the phase variations are very small and will have negligible impact on the station performance.

Similar to previous results from AAVS1, the phase calibration solutions were found to be linear with frequency, indicating that calibration phase in each antenna is due entirely to uncorrected differential delay. These delays are corrected to within a sample in the TPMs.
Following from AAVS1, daily calibration scans solved for the differential delays between antennas, and these delays are also stable on timescales of months.
Based on these results, calibration solutions from daily solar calibration scans were applied to data for the entire day for imaging and station beamforming.
These calibration solutions were scaled in amplitude based on the sun's position in the X and Y polarization antenna beam pattern, hence the absolute flux scale of the data is set by the known flux density of the quiet sun\cite{2009LanB...4B..103B} and the dipole beam power patterns in X and Y\cite{9040892}.

\begin{figure}
\includegraphics[width=\linewidth]{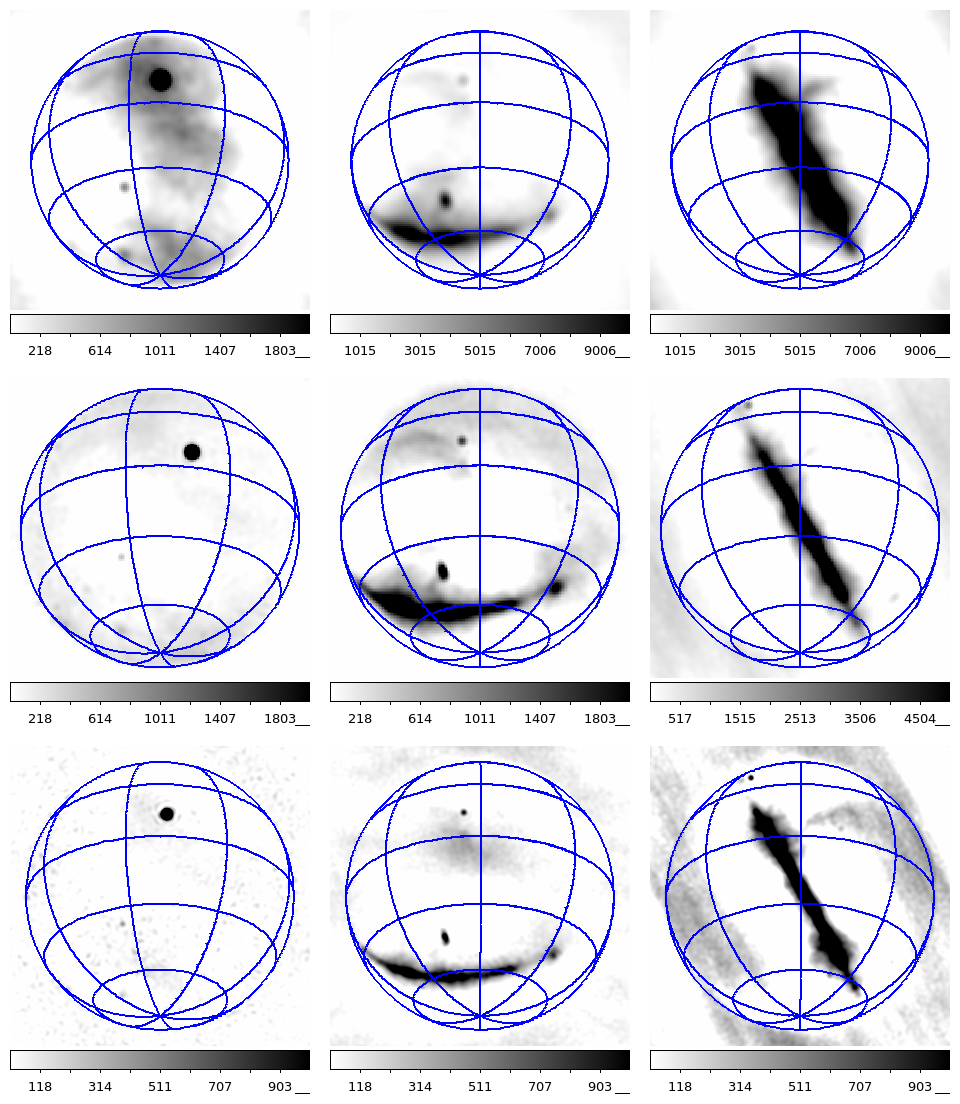}
\caption{Example snapshot sky images. Rows (top to bottom) correspond to images from 110, 159 and 230\,MHz respectively. Columns (left to right) correspond to LSTs of approximately 2, 12 and 18 hours respectively. Note the colorbars have different scale due to the large variation in sky brightness at different LSTs and frequencies. The colorbar is in Jy/beam.}
\label{fig:example_allsky}
\end{figure}

Example snapshot all-sky images are shown in Fig. \ref{fig:example_allsky} for three different frequencies (110, 159 and 230\,MHz) at three different local sidereal times (LSTs; 2,12 and 18\,hours). The LST 02 images include the sun and demonstrate that it was the dominant source in the sky at time of calibration.

\subsection{Temperature dependent calibration amplitude variations}
\label{sec:amp_cal}

A significant outcome of the EDA2 and AAVS2 prototype arrays was the discovery of temperature dependent amplitude variation of the optical signals generated by the FEMs. Since the RFoF system uses optical intensity modulation, any change in absolute signal power from the laser translates to a multiplicative amplitude gain variation in the received signal output from the optical receivers in the TPMs.

The SMART boxes have a similar form factor and solar-reflective powder coat to MWA beamformers, intended to minimize heating during the day from sunlight. However, insolation was found to raise the internal temperatures of the SMART boxes to be approximately 20 degrees above the ambient temperature in the afternoon.
Combined with internally generated heat, the FEMs were operating beyond their design temperature range in the warmer/sunnier months. More analysis and details can be found in a companion paper in these proceedings\cite{Perini2021}.
This issue has been addressed in subsequent modifications and design changes for an updated SMART box, however the effect is present in data reported in this paper and was addressed in post-processing.
Because a single calibration solution was applied to all data for a day, the gain amplitude variations have the effect of changing the absolute flux scale of the data without affecting image quality.

\begin{figure}
\includegraphics[width=\linewidth]{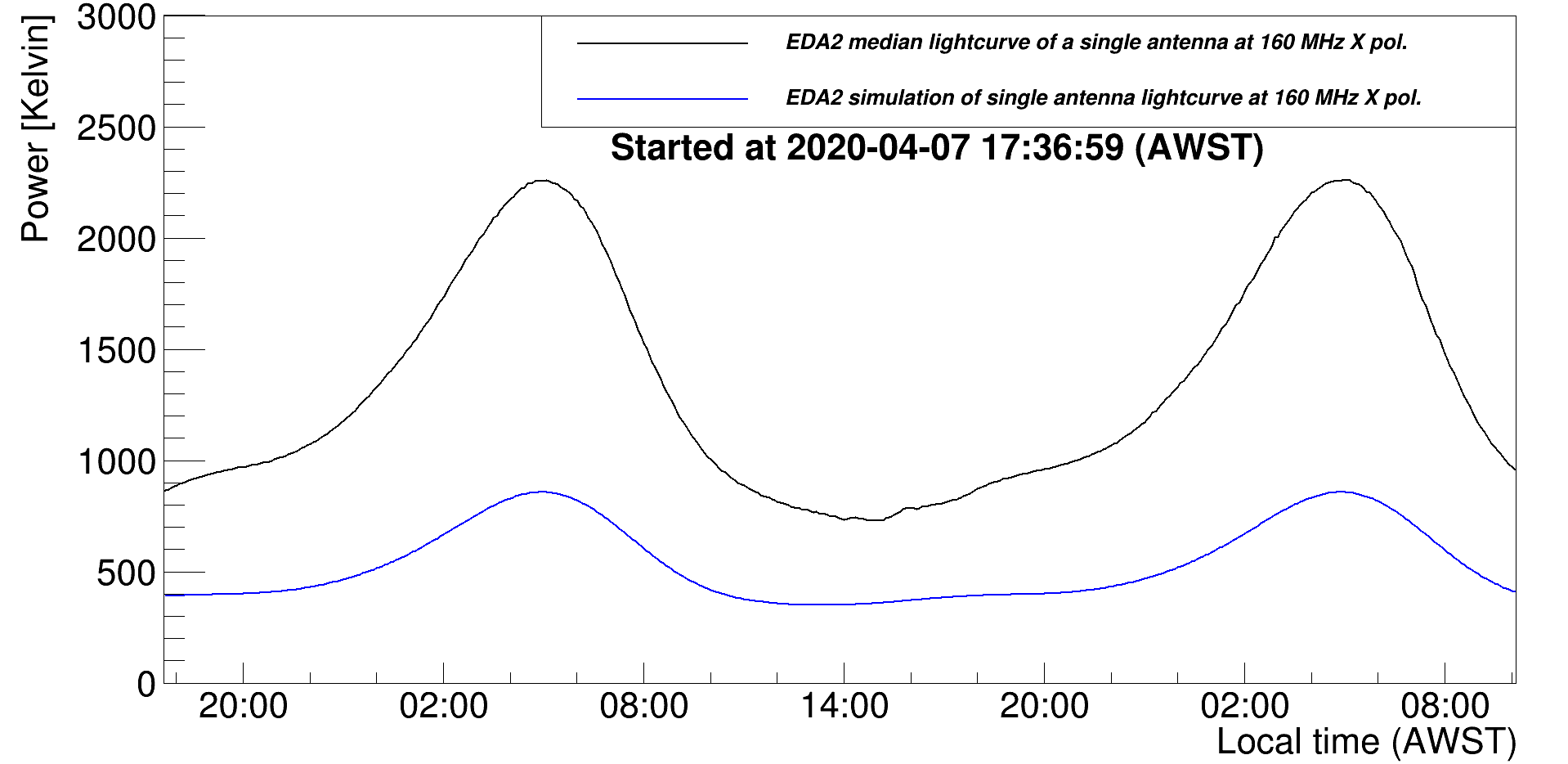}
\caption{Median total power vs time over all antennas vs a model predicted from the average dipole beam pattern and the Haslam all-sky map scaled to 159\,MHz. These are used to calculate a time dependent amplitude gain. Note the simulated data are in Kelvin, but the real data are uncalibrated and scaled for clarity.}
\label{fig:model_vs_median_ant_temp}
\end{figure}

To measure and correct for the gain amplitude variations, the total power as a function of LST, $P_{a,p}(\nu,\text{LST})$, was calculated for each antenna ($a$) and polarization ($p$) using the auto-correlations from the off-line correlated voltage-dumps. 
The expected total power, $T_{sys}(\nu,\text{LST})$, was calculated following a previously used procedure \cite{9410962} (their equation 6), which will be briefly summarised here.
$T_{sys}(\nu,\text{LST})$, the total system temperature, is the sum of receiver temperature $T_{rcv}$ (assumed to be constant in time) and antenna temperature, $T_{ant}(\nu,\text{LST})$, which changes with time due to Earth's rotation.
The model antenna temperature was calculated as a product of model sky temperature $T_{sky}(\nu,\text{LST})$ multiplied by the average response of a dipole beam $B_d(\nu,\theta,\phi)$ in the array and integrated over the entire visible sky.
The model sky temperature $T_{sky}(\nu,\text{LST})$ was calculated using the 408\,MHz Haslam all-sky map\cite{1982A&AS...47....1H}, scaled to the relevant frequency using a spectral index of -2.55.
A median power vs LST dataset, $\hat{P}_{p}(\nu,\text{LST})$, was calculated as the median of $P_{a,p}(\nu,\text{LST})$ over all the antennas;
this and $T_{ant}(\nu,\text{LST})$ are shown in Fig. \ref{fig:model_vs_median_ant_temp}.
The gain, $G_{a,p}(\nu,\text{LST})$, for a given antenna $a$ and polarization $p$ was calculated as $G_{a,p}(\nu,\text{LST}) = P_{a,p}(\nu,\text{LST}) / T_{sys}(\nu,\text{LST})$.
Finally, the final gain as a function of time was calculated as a median of all antenna gain curves (excluding the bad antennas) and normalised to the value at the time of calibration (Sun's transit), resulting in a normalised median gain $\hat{G}_{p}(\nu,\text{LST})$.
The example of $\hat{G}_{p}(\nu,\text{LST})$ at the frequency 159\,MHz MHz in X polarization is shown in Fig. \ref{fig:sens160_gain}.

\added{The amplitude scaling applied by $\hat{G}_{p}(\nu,\text{LST})$ ensures that the absolute flux scale is correct in images formed by the array. This in turn allows the direct measurement of sensitivity as described in the following section.}

\section{Performance}
\label{sec:perf}
\begin{figure}
\includegraphics[width=\linewidth]{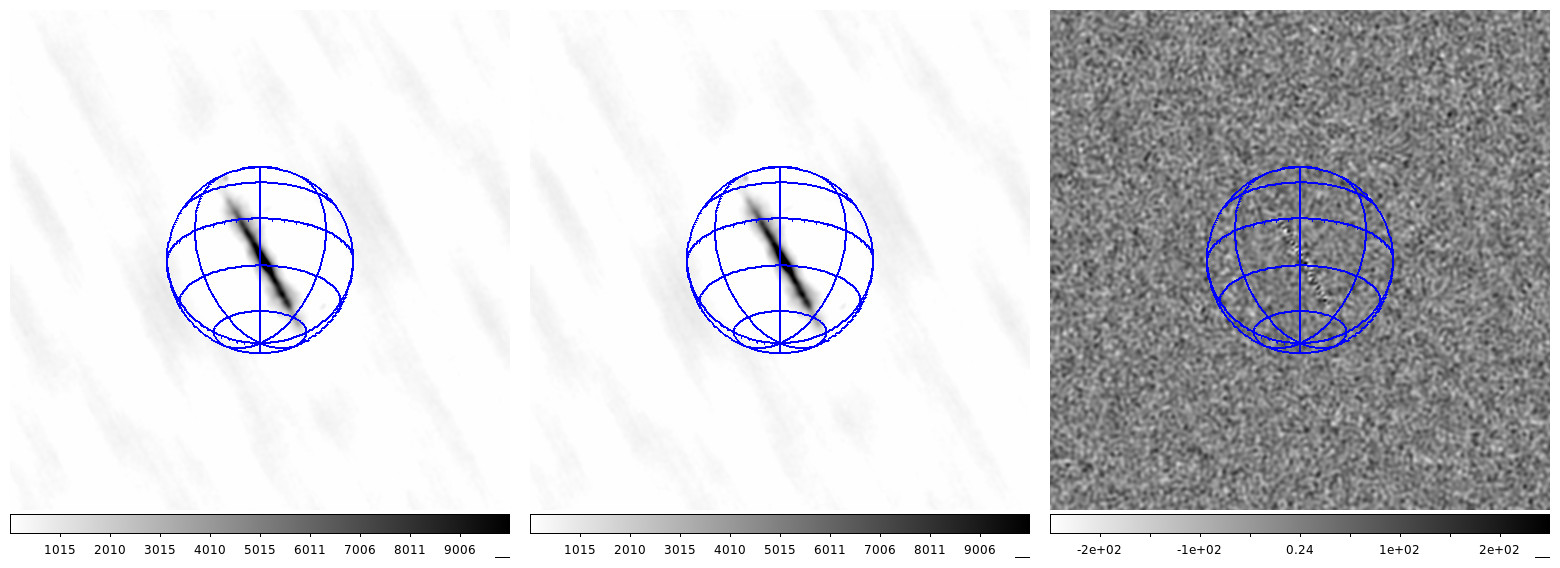}
\caption{Example 159\,MHz sky and difference images used for sensitivity measurements. Left and center: the two consecutive snapshot images. Right: the difference. Note the different flux scales.}
\label{fig:diff_images}
\end{figure}

To measure the sensitivity of the array, we used the same difference imaging technique that was used to measure the sensitivity of AAVS1 and to make a preliminary measurement for AAVS2\cite{BenthemAAVS1,9410962}. We briefly review the technique here.
Each voltage-dump dataset comprises 0.28\,s of data, and the dumps were separated by approximately 5 minutes. Each voltage-dump dataset was correlated offline into two successive 0.14\,s timesteps with zenith as the phase center. A single calibration solution per frequency was applied to all data, and the time-dependent gain amplitude corrections were applied to all data. Each dataset was imaged into two independent 0.14\,s zenith-pointed snapshots that cover the entire hemisphere as shown in Fig. \ref{fig:diff_images}, with each polarization imaged independently.

A difference image was formed by subtracting the second image from the first in each set of two sequential snapshots. Because the rotation of the sky and changes in system calibration are negligible during the short time interval, the difference images are entirely noise: all continuous radio sources and their sidelobes subtract (transient radio sources can be efficiently detected with this technique also\cite{2021PASA...38...23S}).
Example images are shown in Fig. \ref{fig:diff_images}.
Because the images are flux calibrated, the standard deviation of the noise in the difference image is a direct measure of the array's system equivalent flux density (SEFD)\cite{1999ASPC..180..171W}, averaged over all antennas, using the known integration time and bandwidth.
Very strong sources can generate additional source noise at their locations in the map, hence the standard deviation of the difference images is measured in the corners, which is beyond the horizon and cannot contain any actual sources or be affected by source noise.
The SEFD is the inverse of the ``A/T'' metric (effective area divided by system temperature, in $\mathrm{m}^2 \mathrm{K}^{-1}$)  specified by the SKA, specifically $A/T = 2k / \mathrm{SEFD}$ where $k$ is Boltzmann's constant.
The SEFD of a dipole is calculated simply by multiplying the array SEFD into the number of active dipoles, 244 in this case, then converting to A/T.

\begin{figure}
\includegraphics[width=\linewidth]{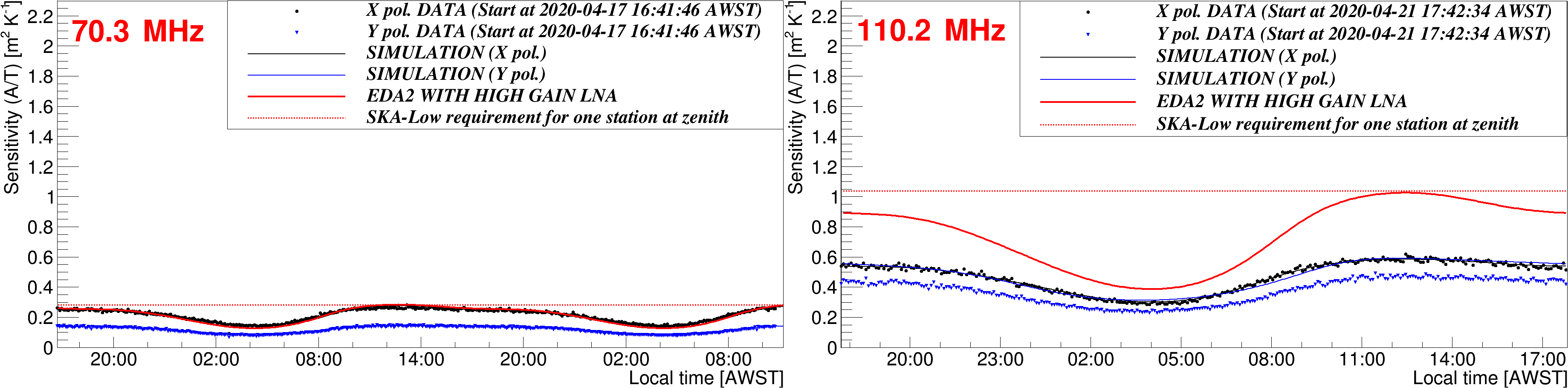}
\caption{Sensitivity: 70 and 110\,MHz. The points show measured on-sky zenith sensitivity for the two polarizations, and the blue and black lines the predicted sensitivity given the EDA2 system temperature. The red dotted straight line shows the SKA1-Low specification sensitivity, which assumes a fixed sky temperature, and the thicker red curve shows the actual EDA2 sensitivity corrected for the increased receiver noise, which would be achieved with a higher gain LNA.}
\label{fig:sens70_110}
\end{figure}
\begin{figure}
\includegraphics[width=\linewidth]{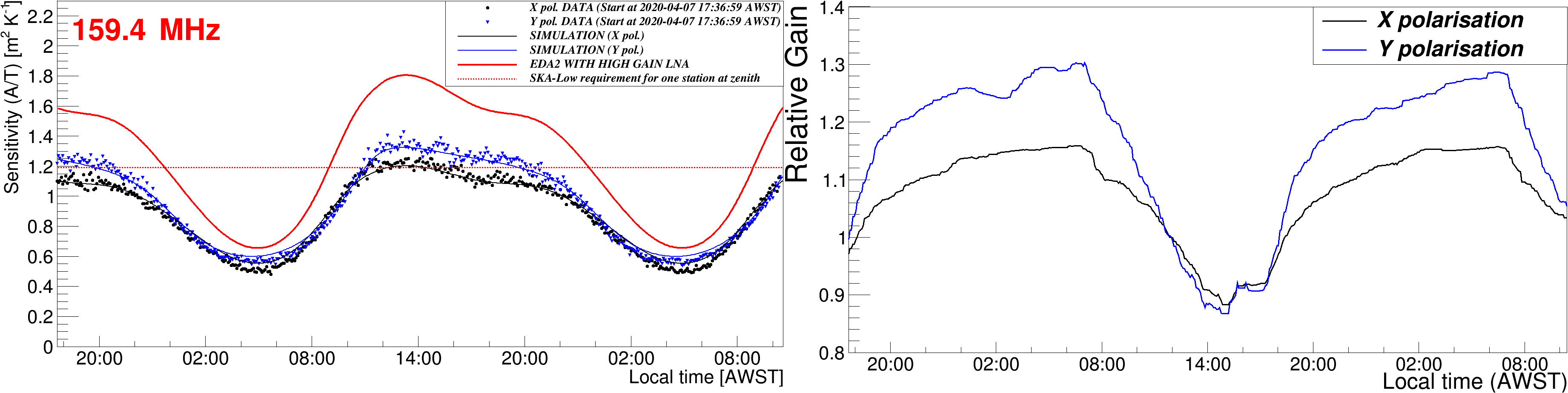}
\caption{Left: Sensitivity at 159\,MHz vs time/LST. Right: The gain amplitude scaling factor applied to the 159\,MHz data to preserve the absolute flux scale.}
\label{fig:sens160_gain}
\end{figure}
\begin{figure}
\includegraphics[width=\linewidth]{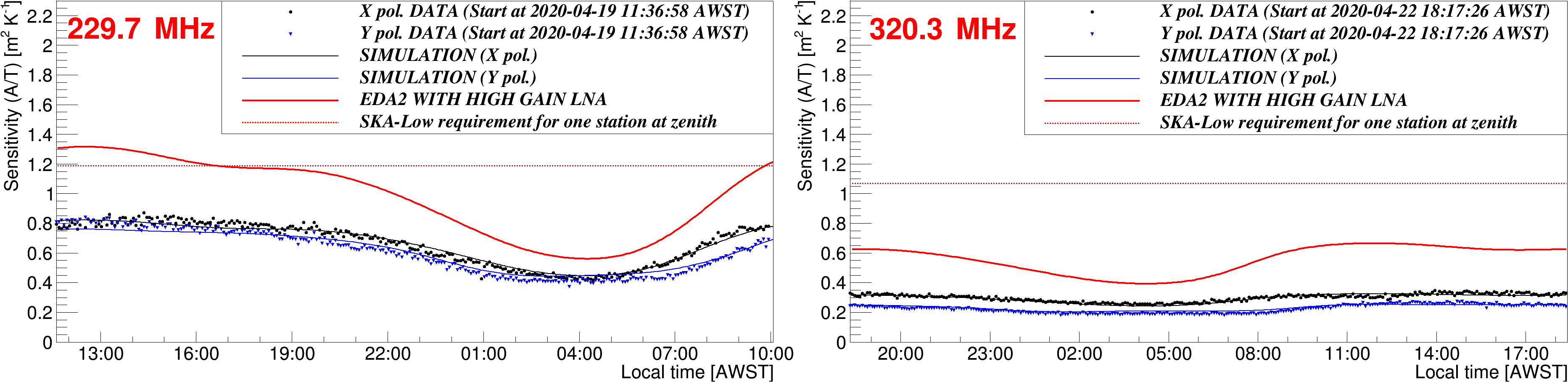}
\caption{Sensitivity at 230 and 320\,MHz. Note that the MWA dipole's peak response at 320\,MHz is not at zenith, however the sensitivity measurements are all referenced to the zenith as \added{previously (pre-2020)} required by the SKA specification. Hence the MWA dipole's sensitivity appears well below specification at this frequency, but a much improved sensitivity would be obtained in reality by simply observing off-zenith at the peak of the dipole's response.}
\label{fig:sens230_320}
\end{figure}

Using this technique, the on-sky zenith-pointed sensitivity of the EDA2, at the five spot frequencies, were calculated and are plotted in Figs \ref{fig:sens70_110}, \ref{fig:sens160_gain}, and \ref{fig:sens230_320} as a function of time.
The measured sensitivity changes as a function of time, because the average sky temperature as seen by each dipole changes during the sidereal and solar day. The extra radio power from the Galactic plane and Galactic center significantly increases the system temperature for a sky-noise dominated system, as does the sun at the higher frequencies. We take this into account when comparing with expected sensitivity in the next section.
In all plots, the minimum in sensitivity was around 4-5\,am local time, corresponding to the Galactic center's transit in April.

\subsection{Expected and specification sensitivity}
\label{sec:sens}

The SKA system requirements\cite{SKA1L1req} \replaced{are independent of LST}{assume a fixed sky temperature}, shown as the dotted red line in Figs \ref{fig:sens70_110}, \ref{fig:sens160_gain}, and \ref{fig:sens230_320}.
The actual on-sky sensitivity depends on the system temperature, which is the sum of the beam-weighted average sky temperature and receiver temperature. The on-sky sensitivity of the original EDA (using identical dipoles and configuration to EDA2) was measured including all effects due to antenna coupling and size at lower frequencies\cite{9040892}. This provides a known reference for the receiver noise.

As noted in section \ref{sec:siglevels}, the EDA2 signal chain suffers from additional noise from the FEMs due to the lower gain LNA.
Given the noise figure of the FEMs, the LNA gain and cable gains, we estimate this to be approximately 70\,K extra receiver noise, but the actual value will be frequency dependent and must be estimated from data. 

\begin{figure}
\includegraphics[width=\linewidth]{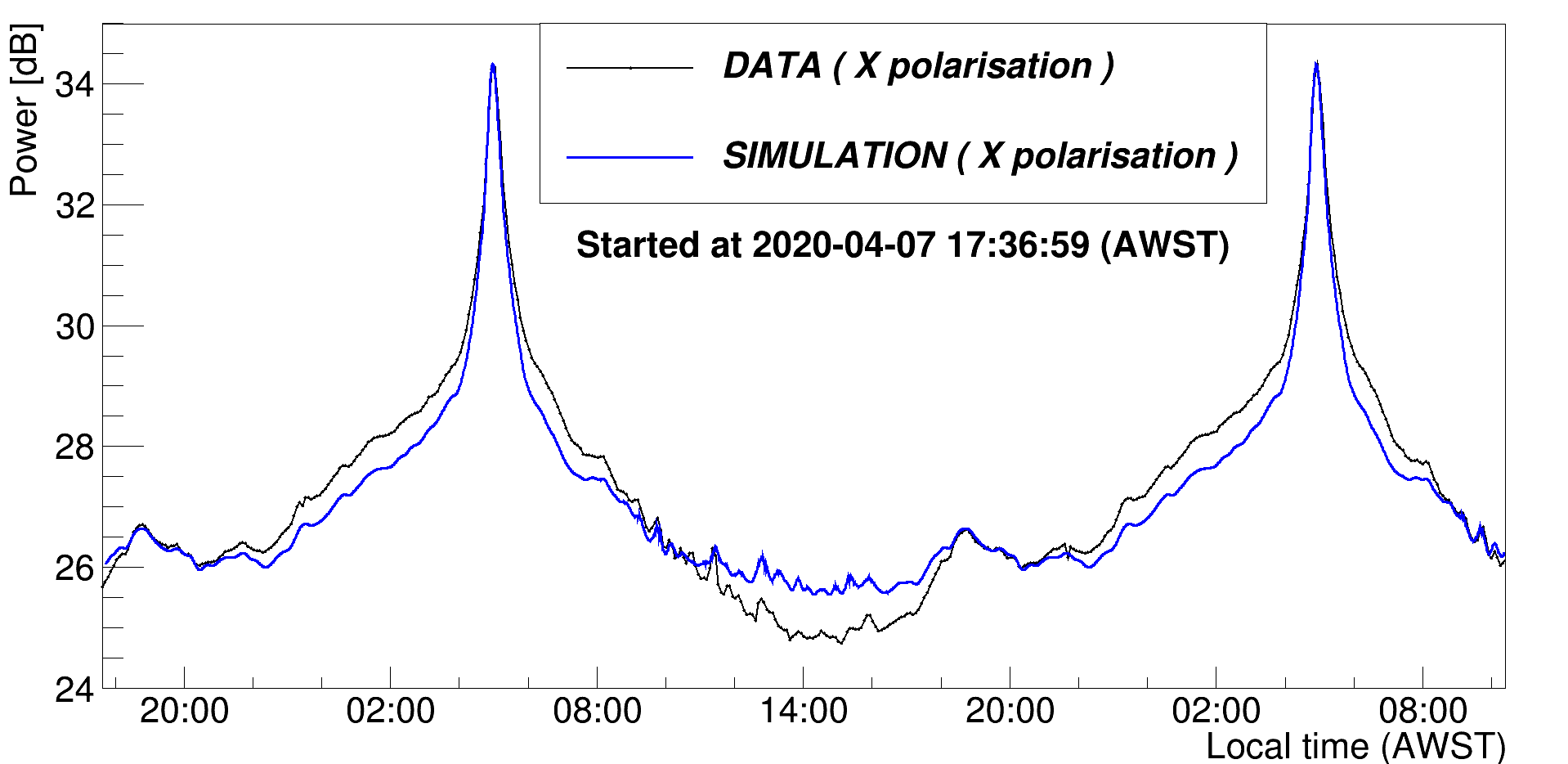}
\caption{Example measured and simulated total-power lightcurves at 159\,MHz from a zenith-pointed array beam.}
\label{fig:lightcurve160}
\end{figure}

We measured the total receiver temperature of the EDA2 using the same technique as was used for the original EDA\cite{2017PASA...34...34W}, which we briefly review here. The array is configured to form a zenith station beam. This produces a characteristic total power lightcurve over a day, with a significant peak as the Galactic center passes overhead.
We predict the received power from the array using the Haslam 408\,MHz all-sky map\cite{1982A&AS...47....1H} scaled to the relevant frequency (using the same technique as above), a model of the dipole power pattern (including effective area), and the array factor using the known locations of the antennas.
This produces a model lightcurve of received power (represented by the antenna temperature $T_{ant}$) over a day. An example of the predicted and measured total power lightcurve of the array is shown in Fig \ref{fig:lightcurve160}.
We note that this procedure is different to the procedure to solve for temperature dependent gains in section \ref{sec:amp_cal}. Here, a single array beam is being compared to a model array beam, and the array factor is important. In section \ref{sec:amp_cal}, the median total power calculated over each dipole was being compared to a model of the power seen by a single dipole.

The measured power is given by
    $P=g(T_{ant} + T_{rec})$
where the (frequency dependent) gain, $g$, and receiver temperature, $T_{rec}$, are unknown, assumed constant.
The two unknowns are found by fitting the data to the model over the 24-hour drift scan.
This process measured the total receiver temperature of the EDA2 at each spot frequency, and we attribute excess noise compared to the original EDA to the extra noise in the FEMs that has not been hidden by a higher gain LNA.

The blue and black curves in Figs \ref{fig:sens70_110}, \ref{fig:sens160_gain}, and \ref{fig:sens230_320} show the \emph{predicted} sensitivity of the EDA2 incorporating this extra noise, and the agreement between model and data is excellent. We reiterate that the actual sensitivity measurements were generated with a completely independent technique that makes no assumptions about the sky or receiver temperature, and is set by the known flux scale of the quiet sun.

Finally, the solid red curve in Figs \ref{fig:sens70_110}, \ref{fig:sens160_gain}, and \ref{fig:sens230_320} show the sensitivity of the EDA2 corrected for the additional noise from the FEMs. This curve  is labelled ``High Gain LNA'', since it represents what the measured sensitivity would be if the noise of the FEMs was hidden with a higher gain LNA. This curve shows the true underlying sensitivity of the EDA2 with MWA dipoles. The dipoles perform well above specification at the spot frequencies of 159 and 230\,MHz, which are within the optimum design range of the dipoles (approximately 140-230\,MHz). At lower frequencies the dipole just meets specification (70 and 110\,MHz) at a few LSTs.
This is to be expected, because the MWA dipoles are not working in their optimum range at the lower frequencies and suffer from a poor antenna match\cite{9040892}.
The 320\,MHz data is noteworthy because at first glance it appears the dipole is well below \added{zenith-directed} specification, \deleted{but the SKA specification is for zenith-directed sensitivity} however the peak of the dipole response is off zenith at higher frequencies. To be consistent with the SKA requirements, we have measured the sensitivity relative to zenith, but reiterate that it is actually not the appropriate measure when the peak directivity of the antenna is off zenith.

\section{Conclusion}
\label{sec:conclusion}
The EDA2 was deployed along with AAVS2 as the second generation of SKA-Low prototype arrays. The EDA2 was deployed as a comparator array in the bridging phase of the SKA, using a known and well understood bowtie dipole antenna that has low mutual coupling.
While retaining the overall signal chain architecture of the previous generation system, EDA2 introduced some significant equipment changes including the use of SMART boxes and the FNDH to address practicality issues with  deployment and maintenance.

The array was used to verify intra-day and longer timescale calibration phase stability, and to directly measure the on-sky sensitivity of the array. These measurements discovered a gain amplitude variation in the optical modules caused by excess temperature, which will be addressed in future designs. The array was used to show these variations are of order $\pm 10\%$ over a day.

The station sensitivity was determined by using the array as an imaging interferometer and measuring the station SEFD directly from noise-dominated difference images. This technique makes no assumptions about the system temperature and only depends on the known flux density of the quiet Sun. A model for the predicted sensitivity of the array, including excess system temperature from lower gain dipoles, agrees very well with the measured data. The sensitivity measurements demonstrate that the dipole performs well above SKA specification at 159 and 230\,MHz and meets specification at 70 and 110\,MHz. At 320\,MHz the peak gain of the dipole is off-zenith \replaced{hence appears lower than the zenith-directed sensitivity specification.}{whereas the SKA specification is for zenith, rather than in the conventional direction of peak directivity}.

The EDA2 has thus demonstrated that a practical and sensitive station can be constructed using simple bowtie dipole antennas in an otherwise identical SKA1-Low signal chain.
\added{The dipole is not as sensitive as the baseline design log periodic antenna over the full frequency range for SKA-Low, but} as a comparator system, the EDA2 presents a viable model in a different region of the performance--cost--risk parameter space of system design.

\acknowledgments
AAVS2 and EDA2 are hosted by the MWA under an agreement via the MWA External Instruments Policy.
This scientific work makes use of the Murchison Radio-astronomy Observatory, operated by CSIRO. We acknowledge the Wajarri Yamatji people as the traditional owners of the Observatory site.
This work was further supported by resources provided by the Pawsey Supercomputing Centre with funding from the Australian Government and the Government of Western Australia. The acquisition system was designed and purchased by INAF/Oxford University and the RX chain was design by INAF, as part of the SKA design and prototyping program.


\newcommand{\pasa}{PASA}
\newcommand{\aap}{A\&A}
\newcommand{\aj}{AJ}
\newcommand{\mnras}{MNRAS}
\bibliography{refs}   
\bibliographystyle{spiejour}   


\vspace{2ex}\noindent\textbf{Wayth} is an Associate Professor at the Curtin University node of the International Centre for Radio Astronomy Research (ICRAR). He received Bachelors degrees in Electrical Engineering and Computer Science from the University of Melbourne in 1995, and his PhD degree in astrophysics from the University of Melbourne in 2005.

\vspace{1ex}
\noindent Biographies and photographs of the other authors are not available.

\listoffigures
\listoftables

\end{spacing}
\end{document}